\documentclass[%
 print,
superscriptaddress,
 amsmath,amssymb,
rmp,
]{revtex4-2}

\usepackage[utf8]{inputenc}
\usepackage[T1]{fontenc}
\usepackage{mathptmx}
\usepackage{subfig}
\usepackage{tabularx}
\usepackage{booktabs}
\usepackage{float}
\usepackage[italian,english]{babel}
\usepackage[dvipsnames]{xcolor}
\usepackage[breaklinks,pdftex,pdfpagelabels,bookmarks]{hyperref}
\hypersetup{colorlinks=true,linktocpage=true,pdfstartpage=3,pdfstartview=FitH,breaklinks=true,pdfpagemode=UseNone, pageanchor=true,pdfpagemode=UseOutlines,plainpages=false,bookmarksnumbered,bookmarksopen=true,bookmarksopenlevel=1,hypertexnames=true,pdfhighlight=/0,urlcolor=Brown,linkcolor=MidnightBlue!60!Black,citecolor=PineGreen}
\usepackage[version=4]{mhchem}
\usepackage{physics}
\usepackage{siunitx}
\usepackage{comment}
\usepackage{graphicx}
\usepackage{dcolumn}
\usepackage{bm}
\usepackage{upgreek}
\usepackage{xcolor}
\usepackage{lineno,blindtext}

\newcommand{\ped}[1]{_{\text{#1}}}

\setcitestyle{super,comma,sort&compress}

\begin{document}

\preprint{APS/123-QED}

\author{Halima Giovanna Ahmad*}
\affiliation{Dipartimento di Fisica E. Pancini$,$ Università degli Studi di Napoli Federico II$,$ Monte S. Angelo$,$ via Cinthia$,$ I-80126 Napoli$,$ Italy}
\affiliation{Seeqc$,$ Strada Vicinale Cupa Cinthia$,$ 21$,$ I-80126 Napoli$,$ Italy}
\affiliation{CNR-SPIN$,$ UOS Napoli$,$ Monte S. Angelo$,$ via Cinthia$,$ I-80126 Napoli$,$ Italy}
\author{Martina Minutillo}
\affiliation{Dipartimento di Fisica E. Pancini$,$ Università degli Studi di Napoli Federico II$,$ Monte S. Angelo$,$ via Cinthia$,$ I-80126 Napoli$,$ Italy}
\affiliation{CNR-SPIN$,$ UOS Napoli$,$ Monte S. Angelo$,$ via Cinthia$,$ I-80126 Napoli$,$ Italy}
\author{Roberto Capecelatro}
\affiliation{Dipartimento di Fisica E. Pancini$,$ Università degli Studi di Napoli Federico II$,$ Monte S. Angelo$,$ via Cinthia$,$ I-80126 Napoli$,$ Italy}
\author{Avradeep Pal}
\affiliation{Department of Materials Science and Metallurgy$,$ University of Cambridge$,$ 27 Charles Babbage Road$,$ Cambridge CB3 0FS$,$ United Kingdom}
\affiliation{Department of Metallurgical Engineering and Materials Science$,$ IIT Bombay$,$ Mumbai$,$ 400076$,$ India}
\author{Roberta Caruso}
\affiliation{Dipartimento di Fisica E. Pancini$,$ Università degli Studi di Napoli Federico II$,$ Monte S. Angelo$,$ via Cinthia$,$ I-80126 Napoli$,$ Italy}
\affiliation{Brookhaven National Laboratory$,$ Upton$,$ New York 11973-5000$,$ USA}
\author{Gianluca Passarelli}
\affiliation{Dipartimento di Fisica E. Pancini$,$ Università degli Studi di Napoli Federico II$,$ Monte S. Angelo$,$ via Cinthia$,$ I-80126 Napoli$,$ Italy}
\affiliation{CNR-SPIN$,$ UOS Napoli$,$ Monte S. Angelo$,$ via Cinthia$,$ I-80126 Napoli$,$ Italy}
\author{Mark Blamire}
\affiliation{Department of Materials Science and Metallurgy$,$ University of Cambridge$,$ 27 Charles Babbage Road$,$ Cambridge CB3 0FS$,$ United Kingdom}
\author{Francesco Tafuri}
\affiliation{Dipartimento di Fisica E. Pancini$,$ Università degli Studi di Napoli Federico II$,$ Monte S. Angelo$,$ via Cinthia$,$ I-80126 Napoli$,$ Italy}
\author{Procolo Lucignano}
\affiliation{Dipartimento di Fisica E. Pancini$,$ Università degli Studi di Napoli Federico II$,$ Monte S. Angelo$,$ via Cinthia$,$ I-80126 Napoli$,$ Italy}
\affiliation{CNR-SPIN$,$ UOS Napoli$,$ Monte S. Angelo$,$ via Cinthia$,$ I-80126 Napoli$,$ Italy}
\author{Davide Massarotti§}
\affiliation{Dipartimento di Ingegneria Elettrica e delle Tecnologie dell’Informazione$,$ Università degli Studi di Napoli Federico II$,$ via Claudio$,$ I-80125 Napoli$,$ Italy}
\affiliation{CNR-SPIN$,$ UOS Napoli$,$ Monte S. Angelo$,$ via Cinthia$,$ I-80126 Napoli$,$ Italy}

\title{Coexistence and tuning of spin-singlet and triplet transport in spin-filter Josephson junctions}

\maketitle

*halimagiovanna.ahmad@unina.it

§davide.massarotti@unina.it

	\section*{abstract}
	The increased capabilities of coupling more and more materials through functional interfaces are paving the way to a series of exciting experiments and extremely advanced devices. Here we focus on the capability of magnetically inhomogeneous superconductor/ferromagnet (S/F) interfaces to generate spin-polarized triplet pairs. We build on previous achievements on spin-filter ferromagnetic Josephson junctions (JJs) and find unique correspondence between neat experimental benchmarks in the temperature behavior of the critical current and theoretical modelling based on microscopic calculations, which allow to determine \emph{a posteriori} spin-singlet and triplet correlation functions. This kind of combined analysis provides an accurate proof of the coexistence and tunability of singlet and triplet transport. This turns to be an opportunity to model disorder and spin-mixing effects in a JJ to enlarge the space of parameters, which regulate the phenomenology of the Josephson effect and could be applied to a variety of hybrid JJs.

\section{Introduction}

The growing complexity of the Josephson junctions (JJs) in terms of layout and materials has significantly increased the "parameters space" for a full understanding and control of their properties~\cite{Barone1982,Tafuri2019}. Ferromagnetic (SFS) JJs are a unique platform to integrate the coherent quantum nature of superconductors and ferromagnets into unconventional mechanisms and smart tunable functionalities. The rich literature has established several key elements, which arise when superconducting pair correlations traverse the exchange field of a ferromagnet~\cite{Ryazanov2001,Golubov2004,Buzdin2005,Bergeret2005,Blamire2014,Peng2017,Zhang2020}. JJs with multiple F-layer barriers have been theoretically and experimentally studied in connection to unconventional triplet superconductivity with equal-spin Cooper pairs, characterized by total spin momentum $S=1$ and spin $z$-component $S_z=\pm 1$ ($\ket{11}_{S,S_z}=\ket{\uparrow\uparrow}$ and $\ket{1-1}_{S,S_z}=\ket{\downarrow\downarrow}$), which can be artificially generated in these structures~\cite{Buzdin2005,Bergeret2005,Robinson2010,Khaire2010,Sprungmann2010,Eschrig2011,Banerjee2014,Linder2015,Singh2015,Eschrig2015,Martinez2016}. Compared to spin-singlet Cooper pairs and opposite-spin triplet Cooper pairs (total spin momentum $S=1$ and spin $z$-component $S_z=0$, $\ket{10}_{S,S_z}=1/\sqrt{2}(\ket{\uparrow\downarrow}+\ket{\downarrow\uparrow})$), the spin-aligned triplet Cooper pairs are immune to the exchange field of the F layer and can carry a non-dissipative spin current. Therefore, spin-triplet Cooper pairs constitute the essential element for the emerging field of superconducting spintronics~\cite{Robinson2010,Khaire2010,Eschrig2011,Linder2015,Eschrig2015,Martinez2016}.

It is well established that spin-polarized triplet pairs are generated via spin-mixing and spin-rotation processes at magnetically inhomogeneous S/F interfaces~\cite{Bergeret2005,Eschrig2011,Eschrig2015,Linder2015,Houzet2007,Cascales2019}. Evidence of equal-spin triplets has been reported in S-F'-F-F''-S JJs, where the F', F'' spin-mixer layers mediate the conversion of singlet to triplet pair correlations~\cite{Robinson2010,Khaire2010,Sprungmann2010,Banerjee2014,Iovan2014,Linder2015,Singh2015,Martinez2016,Massarotti2018}. Recently, theoretical and experimental studies have been dedicated to an alternative mechanism for triplet pair generation involving spin-orbit coupling (SOC) in combination with a magnetic exchange field~\cite{Edelshtein1989,Banerjee2018,KunRok2020}. These systems may benefit from the capability to generate controllable spin-polarized supercurrents with a single ferromagnetic layer, compared to magnetically textured JJs.

The strong evidence for the presence of spin-triplet supercurrents in a JJ is the slower decay of the characteristic voltage of the junction with increasing F layer thickness~\cite{Robinson2010,Khaire2010,Sprungmann2010,Eschrig2011,Anwar2012,Blamire2014,Linder2015,Eschrig2015,Singh2015,Martinez2016}, due to the robustness of spin-triplet Cooper pairs to the exchange field. However, it has been suggested that a mechanism of phase-compensation can arise in clean S/F heterostructures, which may cancel the destructive interference effect due to the exchange field on conventional spin-singlet pairing~\cite{Melnikov2012,Iovan2014}. Therefore, a conclusive evidence for the spin-triplet nature of the supercurrent could be supported by the capability to distinguish singlet and triplet components. The capability of quantifying the amount of spin-polarized supercurrents remains a fundamental benchmark to further prove triplet correlations and a key step towards real applications.

In parallel with the work on diffusive ferromagnets, superconducting tunnel junctions with ferromagnetic insulator (FI) barriers (SFIS), namely spin-filter \ce{NbN}/\ce{GdN}/\ce{NbN} JJs, have revealed unique transport properties, such as spin-polarization phenomena~\cite{Senapati2011,28,Pal2014,Ahmad2020JOSC}, an interfacial exchange field in the superconducting layer~\cite{Pal2015,Zhu2016}, macroscopic quantum tunneling~\cite{Massarotti2015} and an unconventional incipient $0$-$\uppi$ transition~\cite{Caruso2019}. They are especially well-suited for the implementation in superconducting circuits in which a very low dissipation is required~\cite{Bannykh2009,Wild2010,Kawabata2006,Kawabata2010,Feofanov2010,Zhu2017,Ahmad2020}. In these systems, evidence of spin-triplet transport has been reported~\cite{Pal2014,Blamire2014,Pal2017,Caruso2019}. 

Here we build on a previous study of the critical current $I\ped c$ as a function of the temperature $T$ in \ce{NbN}-\ce{GdN}-\ce{NbN} JJs~\cite{Caruso2019}, now performed in presence of magnetic field to demonstrate coexistence and tuning of singlet and triplet components. By using a tight-binding Bogolioubov de Gennes (BdG) approach~\cite{Furusaki1994, Asano2019, Asano2001,minutillo2021realization}, we model the $I\ped c(T)$ curves in the whole temperature range, along with the corresponding current-phase relation (CPR) as a function of the temperature $T$. It turns out that measurements of the temperature behavior of the critical current along with microscopic modelling approach provide an alternative accurate method to assess the spin-triplet transport, which can be extended to different types of JJs: the amount of spin-singlet and -triplet correlations can be quantified and parametrized in terms of disorder parameter and spin-mixing mechanisms through a direct fitting of experimental data. 

The large variety of materials and configurations employed in diffusive SFS JJs in literature allows to access a wide range of behaviors for the thermal dependence of the $I\ped c$~\cite{Barone1982,Buzdin2005}. Particularly relevant to our work, the possibility of generating oscillations in the superconducting order parameter by means of a finite exchange field in the F interlayer results in a $\uppi$ phase-shift in the CPR and a sudden drop of the $I\ped c$ towards zero at temperatures $T_{\uppi}<T\ped c$, followed by an increase of the $I\ped c$ for $T>T_{\uppi}$~\cite{Ryazanov2001,Buzdin2005}. Therefore, the non-monotonic behavior for the $I\ped c(T)$ in systems in which a $0$-$\uppi$ transition occurs is characterized by a peculiar cusp at the transition point $T_{\uppi}$~\cite{Ryazanov2001,Buzdin2005}. In this work, we focus on the peculiar behavior of the $I\ped c(T)$ in tunnel ferromagnetic spin-filter JJs, in which an unconventional $0$-$\uppi$ transition occurs. Above a \ce{GdN} thickness $d\ped F=\SI{3.0}{\nm}$, the $I\ped c$ is not completely suppressed at $T_{\uppi}$, thus suggesting that a $0$-$\uppi$ transition broadened in a range of temperatures of the order of some $\si{\K}$ occurs~\cite{Caruso2019}. Therefore, in the devices discussed here, the $I\ped c(T)$ curve shows a region in which the $I\ped c$ is constant in a wide range of temperatures, i.e. it shows a plateau, or it shows a non-monotonic trend characterized by a non-zero local minimum, i.e. the $I\ped c(T)$ exhibits an incipient $0$-$\uppi$ transition~\cite{Caruso2019}. Such unconventional $I\ped{c}(T)$ behavior turns out to be the benchmark for the coexistence of spin-singlet and spin-triplet superconductivity in SFIS junctions. When the $I\ped c(T)$ curve shows a plateau over a wide range of temperatures, the competition between the singlet and triplet pairing amplitudes becomes significant, in both s-wave and p-wave symmetries. This behavior sets in due to the combined effects of impurities and spin-mixing mechanisms. When the $I\ped{c}(T)$ curve exhibits an incipient $0-\uppi$ transition, the equal-spin triplet component is gradually suppressed, becoming irrelevant in the limit case of a more standard cusp-like $0-\uppi$ transition. This last situation corresponds to relative low values of disorder and spin-mixing effects. 

An external magnetic field perpendicular to the Josephson transport direction gradually modifies the experimental $I\ped c(T)$ curves, i.e. a plateau extended over a wide range of temperatures evolves into a non-monotonic behavior by increasing the magnetic field. To the best of our knowledge, this represents the experimental evidence of an "in situ" tuning of the relative weight between spin-singlet and spin-triplet supercurrents in single-layered SFIS JJs, which is explained in terms of a reduced disorder parameter in presence of magnetic field for a multi-domain ferromagnet.  The ability to describe the combined effect of magnetic inhomogeneities and disorder in complex barriers, with clear benchmarks on the phenomenology of the junctions, can be of reference for a variety of structures.

\section{Results}

\subsection{Spin-filter Josephson junctions and microscopic modelling}

\begin{figure}[t]
	\centering 
	\includegraphics[width=0.7\columnwidth]{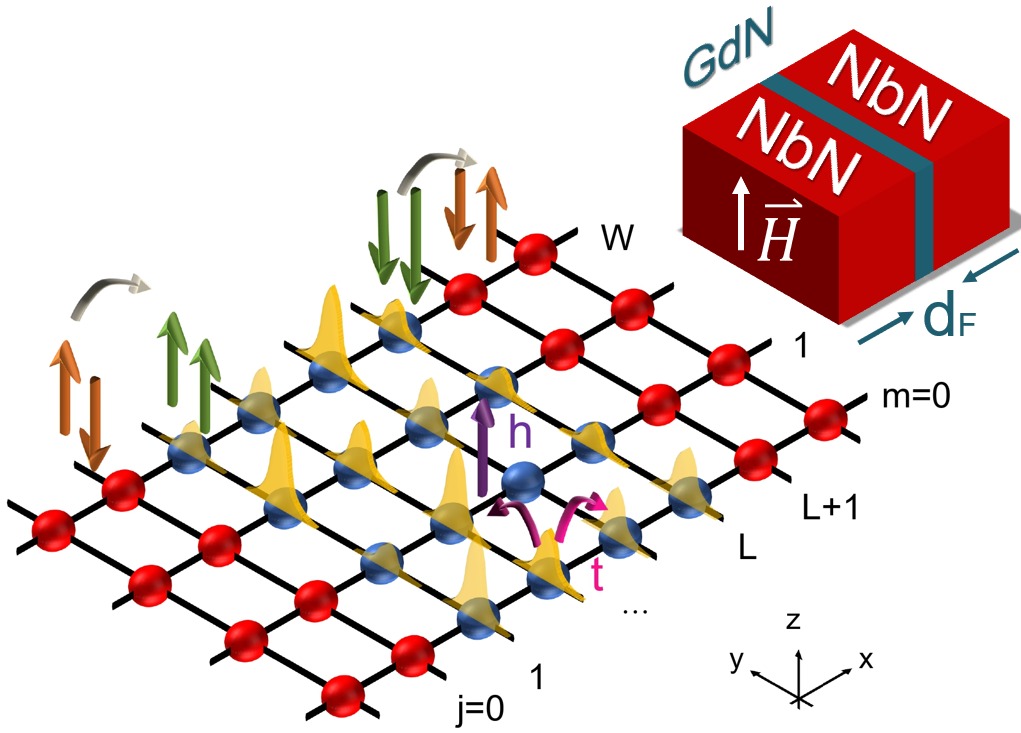}
	\caption[]{\textbf{Spin-filter Josephson junction and two-dimensional lattice model scheme}. Picture of the Superconductor/Insulating ferromagnet/Superconductor two-dimensional lattice model. The barrier (highlighted in blue) has a total thickness $L$ along $\mathbf{x}$. The junction width is $W$ along $\mathbf{y}$. The spin-mixing mechanism due to the spin-orbit coupling is depicted by the spin-flipping process highlighted at the interface between the superconducting boundaries (red sites) and the barrier. The impurities, with random strength depicted by the height of the yellow potential peaks, are represented on each site of the lattice. The exchange field $h$ (violet arrow) is parallel to the $\mathbf{z}$ axis, while the hopping $t$ between nearest-neighbour sites is here represented by pink arrows. In the inset, sketch of \ce{NbN}-\ce{GdN}-\ce{NbN} Josephson junctions reported in this work. The external magnetic field $H$ is parallel to the $\mathbf{z}$ axis.}
	\label{Fig_4}
\end{figure}

The junctions under study are \ce{NbN}/\ce{GdN}/\ce{NbN} JJs~\cite{Caruso2019}, with a special focus on devices with thick FI layers. The superconducting electrodes have thicknesses of $\SI{100}{\nm}$ and the analyzed FI thicknesses are: $d\ped F=\SI{3.0}{\nm}$, $\SI{3.5}{\nm}$ and $\SI{4.0}{\nm}$. A sketch of the JJs is reported in the inset of Figure~\ref{Fig_4}. In this work, we report measurements of the $I\ped c(T)$ curves performed down to $\SI{20}{\milli\K}$ and measurements of the $I\ped c(T)$ curves as a function of an external magnetic field.

We model the S/FI/S junctions using a tight-binding BdG Hamiltonian on a two-dimensional (2D) lattice~\cite{Madelung2012,Ashcroft1976}. A schematization of the 2D-lattice model is reported in Figure~\ref{Fig_4}, where $L$ is the length of the FI barrier and $W$ is the width of the junction expressed in lattice units.

The Hamiltonian of the junction in the Nambu$\otimes$spin space is given by~\cite{Furusaki1994,Asano2001,Asano2019}
\begin{center}
	\begin{equation}
	\label{H_junction}
	\check{\mathcal{H}} = \sum_{\mathbf{r},\mathbf{r}'} \Psi^{\dagger}(\mathbf{r}) \begin{bmatrix} \hat{H}(\mathbf{r},\mathbf{r}') & \hat{\Delta}(\mathbf{r},\mathbf{r}') \\
	-\hat{\Delta}^{*}(\mathbf{r},\mathbf{r}') & - \hat{H}^{*}(\mathbf{r},\mathbf{r}')
	\end{bmatrix} \Psi(\mathbf{r}'),
	\end{equation}
\end{center}
with $	\Psi(\mathbf{r}) = \left[ \psi_{\uparrow}(\mathbf{r}), \psi_{\downarrow}(\mathbf{r}), \psi^{\dagger}_{\uparrow}(\mathbf{r}), \psi^{\dagger}_{\downarrow}(\mathbf{r})  \right]^{T}$. Here, $\psi^{\dagger}_{\mu}(\mathbf{r})$ and $\psi_{\mu}(\mathbf{r})$ are the field operators creating/destructing an electron with spin $\mu$ at the lattice point $\mathbf{r}=j\mathbf{x}+m\mathbf{y}$, with $j=0,1,\dots,L,L+1$ and $m=1,\dots,W$. Here and in the followings, the symbols $\hat{.}$ and $\check{.}$ describe the $2 \times 2$ and $4 \times 4$ matrices, in spin and Nambu$\otimes$spin spaces, respectively.

In Equation~\ref{H_junction}, $\hat{H}$ is the normal-state Hamiltonian of the junction, while $\hat{\Delta}$ describes the superconducting pairing potential. The latter is non-zero only in the S leads, for which a conventional s-wave superconductivity is assumed~\cite{Barone1982}. Thus, $\hat{\Delta}$ is proportional to $\Delta \text{e}^{\text{i}\phi\ped{L}}$ ($\Delta \text{e}^{\text{i}\phi\ped {R}}$) in the left (right) S lead, where $\Delta$ is the order parameter and $\phi\ped{L}$ ($\phi\ped{R}$) is the superconducting phase in the left (right) S lead. The normal-state Hamiltonian $\hat{H}$, instead, can be written as $\hat{H}=\hat{H}\ped{S}+\hat{H}\ped{FI}$, with $\hat{H}\ped{S}$ and $\hat{H}\ped{FI}$ referring to the S-electrodes and the FI barrier, respectively. Their explicit form is shown in the Supplemetary Note 1.

In the S-leads, $\hat{H\ped {S}}$ is described by the parameters $t\ped S$ and $\mu\ped S$, representing the hopping integral among nearest-neighbor lattice sites and the chemical potential, respectively. Relevant parameters of the Hamiltonian $\hat{H\ped {FI}}$, instead, are the hopping integral $t$, the Fermi energy $\mu\ped {FI}$ and the amplitude of the spin-orbit interaction $\alpha$, used to introduce a spin-symmetry breaking~\cite{Madelung2012,Ashcroft1976}. Further, we include on-site random impurity potential with strength $v_{\mathbf{r}}$ uniformly distributed in the range $-V\ped{imp}/2 \leq v_{\mathbf{r}} \leq V\ped{imp}/2$. Finally, an exchange field is assumed to be slightly disordered, and is modeled as $h^{'} = h + \delta_{h}$. Here, $\delta_{h}$ are small on-site fluctuations given randomly in the range $-h/10 \leq \delta_{h} \leq h/10$ (along the $h$-direction). The combined effect of SOC and impurities in the FI region efficiently mimics the presence of magnetic inhomogeneities in the barrier, which are more likely to occur in devices with large areas~\cite{Eschrig2011,Blamire2014}. The junctions under study, in fact, are characterized by areas of $\sim \SI{50}{\micro\m^2}$. This approach is meant to include all possible effects occurring in the FI barrier, and it is a powerful platform to describe a large variety of JJs. As shown below, fitting of experimental $I\ped c(T)$ curves will allow to identify the coexistence of spin-singlet and triplet transport. When compared with what available in literature, the correlation functions are determined a posteriori from the experimental data and allow to quantify the weight between the different transport channels.

The Josephson current $J$ at finite temperature $T$ is derived from the Matsubara Green’s function (GF) of the FI barrier, calculated with the recursive Green's function (RGF) technique~\cite{Furusaki1994,Asano2001,Asano2019,minutillo2021realization}. The barrier Green's function (GF) $\check{G}_{\omega_{n}}(\mathbf{r},\mathbf{r}')$ connecting two lattice sites located at $\mathbf{r}$ and $\mathbf{r}'$ reads
\begin{center}
	\begin{equation}
	\label{Sites_green}
	\check{G}_{\omega_{n}}(\mathbf{r},\mathbf{r}') = \begin{bmatrix}
	\hat{G}_{\omega_{n}} (\mathbf{r},\mathbf{r}') & \hat{F}_{\omega_{n}} (\mathbf{r},\mathbf{r}') \\
	-\hat{F}^{*}_{\omega_{n}} (\mathbf{r},\mathbf{r}') & -\hat{G}^{*}_{\omega_{n}} (\mathbf{r},\mathbf{r}')
	\end{bmatrix} ,
	\end{equation}
\end{center}
and solves the following Gor'kov equation~\cite{Furusaki1994,Asano2001,Asano2019}:
\begin{center}
	\begin{align}
	\label{Gorkov_eq}
	&\left[ \text{i} \omega_{n} \hat{\tau}_{0} \hat{\sigma}_{0} - \sum_{\mathbf{r}_{1}} 
	\begin{pmatrix} 
	\hat{H}(\mathbf{r},\mathbf{r}_{1}) & \hat{\Delta}(\mathbf{r},\mathbf{r}_{1}) \\
	-\hat{\Delta}^{*}(\mathbf{r},\mathbf{r}_{1}) & -\hat{H}^{*}(\mathbf{r},\mathbf{r}_{1})
	\end{pmatrix}   \right] \times \check{G}_{\omega_{n}}(\mathbf{r}_{1}, \mathbf{r}')  = \hat{\tau}_{0} \hat{\sigma}_{0} \delta(\mathbf{r}-\mathbf{r}'),
	\end{align}
\end{center}
Here $\omega_{n}=(2n+1)\uppi T$ is the fermionic Matsubara frequency, $T$ is the temperature and $\hat{\tau}_{0}$ and $\hat{\tau}_{\nu}$ ($\nu=1,2,3$) are the analogous of the identity and Pauli matrices in the Nambu space, respectively. The Josephson current originates from the GFs connecting two adjacent sites along the $\mathbf{x}$-direction (namely $\check{G}_{\omega_{n}}(\mathbf{r},\mathbf{r}+\mathbf{x})$ and $\check{G}_{\omega_{n}}(\mathbf{r}+\mathbf{x},\mathbf{r})$), and reads
	\begin{center}
		\begin{equation}
		\label{Josephson_curr_2}
		J = - \dfrac{\text{i e}}{2} T \sum_{\omega_{n}} \sum_{m=1}^{W} Tr \left[ \hat{\tau}_{3} \check{T}_{+} \check{G}_{\omega_{n}} (\mathbf{r},\mathbf{r} + \mathbf{x}) - \hat{\tau}_{3} \check{T}_{-} \check{G}_{\omega_{n}} (\mathbf{r} + \mathbf{x}, \mathbf{r})   \right] .
		\end{equation}
	\end{center}
Here, $Tr$ stands for the trace over the Nambu$\otimes$spin space and $\check{T}_{\pm}$ matrices describe the hopping and the SOC along the propagation direction (their explicit form is reported in Supplementary Note 1)~\cite{Asano2019}. In order to consider the contribution of the lattice sites along the $\mathbf{y}$-direction, in Equation~\ref{Josephson_curr_2}  we perform a summation $\sum_{m=1}^{W} $. Furthermore, the summation over Matsubara frequencies until convergence is performed. Finally, we calculate the CPR from Equation~\ref{Josephson_curr_2} at fixed temperature $T$ by varying the phase difference $\phi=\phi\ped L-\phi\ped R$ between the S leads from $0$ to $\uppi$. Then, we compute the $I\ped c(T)$ curves from the maximum of the CPRs at $T$ ranging from $0$ to $T\ped c$.

The off-diagonal terms of the matrix in the right-hand side of Equation~\ref{Sites_green} are the so-called anomalous Green’s functions $\hat{F}_{\omega_{n}}$. From these latter, taking the elements with $\mathbf{r}'=\mathbf{r}=j\mathbf{x}+m\mathbf{y}$, we can derive the four pairing components with s-wave symmetry at each position in the barrier ($j=1,\dots,L$) along the $\mathbf{x}$-direction~\cite{Asano2019}: 
\begin{center}
	\begin{equation}
	\label{s_wave_correlations}
	\dfrac{1}{W} \sum_{\omega_{n}}\sum_{m=1}^{W} \hat{F}_{\omega_{n}}(\mathbf{r},\mathbf{r}) = \sum_{\nu=0}^{3} f_{\nu} \hat{\sigma}_{\nu} \; \text{i}  \; \hat{\sigma}_{2} \; ,
	\end{equation}
\end{center}  
where $f_{0}$ is the spin-singlet component, and $f_\nu$, with $\nu=1,2,3$ are the spin-triplet components. In particular, $f_3$ is the opposite-spin triplet component and the equal-spin triplet components $f_{\uparrow}$ ($\;f_{\downarrow}$) are defined as $f_{\uparrow\;(\downarrow)}=if_2\mp f_1$ (see Supplementary Note 1 for details). Here, similarly to Equation(\ref{Josephson_curr_2}), the summation $\sum_{m=1}^{W} $ is performed to take into account all the lattice sites with the same longitudinal coordinate $j$ and different index $m$ in the transverse direction.

Analogous considerations can be applied to the GFs connecting the sites at the position $\mathbf{r}$ with their neighbors in $\mathbf{r}-\mathbf{x}$ and $\mathbf{r}+\mathbf{x}$, from which we can calculate the odd-parity p-wave pairing functions. Thus, the p-wave correlations, at the position $j$ along $\mathbf{x}$ inside the barrier ($\mathbf{r}=j\mathbf{x}+m\mathbf{y}$), can be expressed as:
\begin{gather}
\dfrac{1}{4W} \sum_{\omega_{n}}\sum_{m=1}^{W} \left[\hat{F}_{\omega_{n}}\left( \mathbf{r}+\mathbf{x},\mathbf{r}\right)  + \hat{F}_{\omega_{n}}\left( \mathbf{r},\mathbf{r}-\mathbf{x}\right)  - \hat{F}_{\omega_{n}}\left(\mathbf{r},\mathbf{r}+\mathbf{x}\right)  
-\hat{F}_{\omega_{n}}\left( \mathbf{r}-\mathbf{x},\mathbf{r}\right) \right] = \label{p_wave_correlations} \\ 
\notag\sum_{\nu=0}^{3} f_{\nu} \hat{\sigma}_{\nu} \; \text{i}  \; \hat{\sigma}_{2} .\notag
\end{gather}

\subsection{Experimental results and theoretical interpretation}
\label{results}

\begin{figure}
	\centering
	\includegraphics[width=0.8\columnwidth]{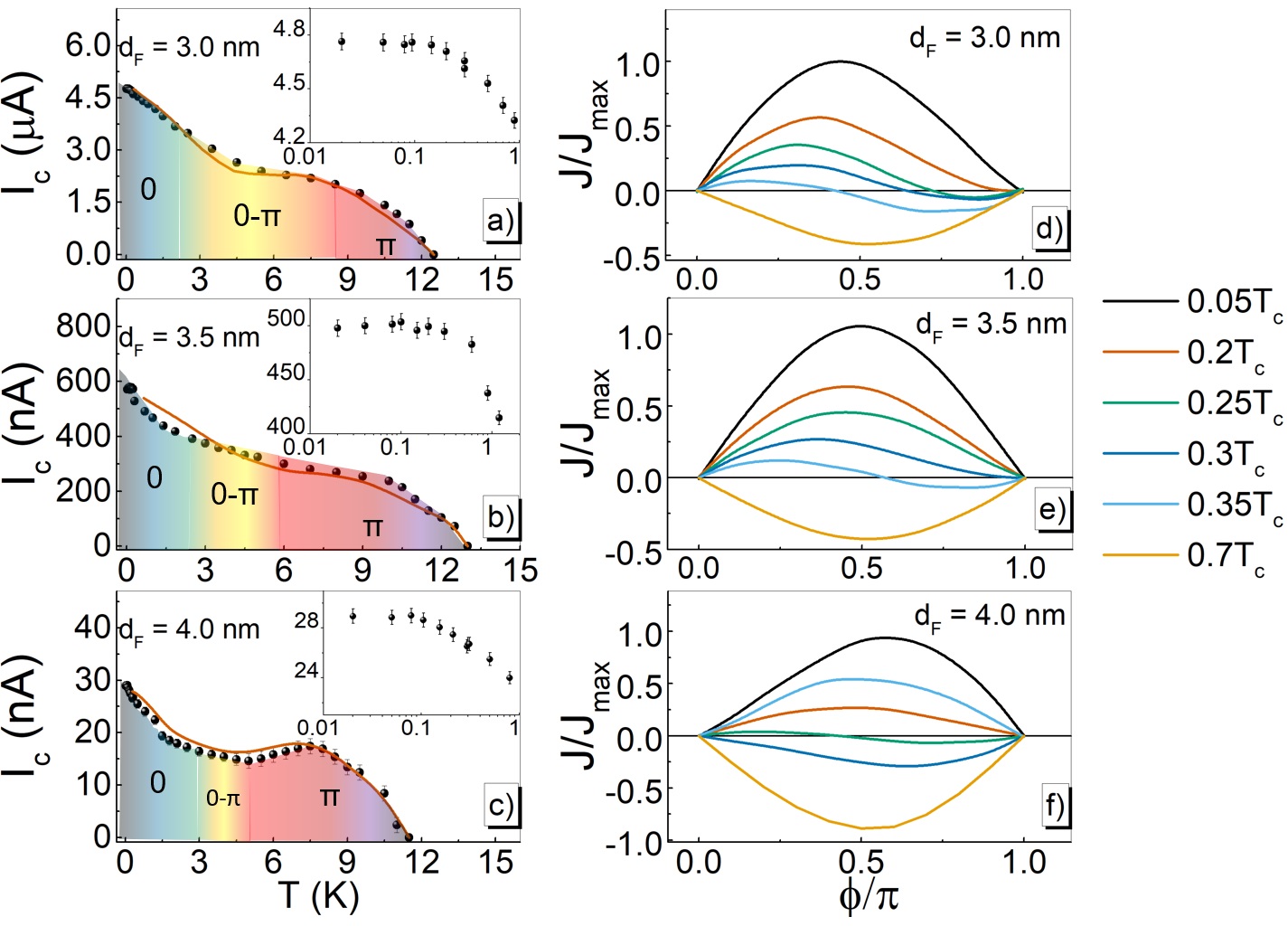}
	\caption[]{\textbf{Comparison between the experimental temperature dependence of the critical current, tight-binding simulations and corresponding current-phase relations.} Critical current $I\ped c$ as a function of the temperature $T$ (black points) for spin-filter junctions with \ce{GdN} barrier thickness $d\ped F=\SI{3.0}{\nm}$ (a),  $d\ped F=\SI{3.5}{\nm}$ (b) and $d\ped F=\SI{4.0}{\nm}$ (c). In the insets of figures (a), (b) and (c): measured saturation of the $I\ped c(T)$ down to $\SI{20}{\milli\K}$. The error bars on the measured $I\ped c$ are of the order of $1\%$ and represent the statistical error due to thermally-induced critical current fluctuations\cite{Massarotti2015}. The red lines are the best $I\ped c(T)$ curves obtained from the maximum of the current-phase relations (CPRs) calculated with the microscopic two-dimensional lattice Bogoliubov-De Gennes  Hamiltonian, with simulations parameters reported in Supplementary Note 2. The amplitude of the simulated critical current has been multiplied by the experimental $I\ped c$ measured at $\SI{20}{\milli\K}$. In (d), (e) and (f): CPRs at selected temperatures near the $0$-$\uppi$ transition to highlight the arising of higher order harmonics, compared with those in the $0$ and $\uppi$ state, at $0.05\;T\ped c$ and $0.7\;T\ped c$, respectively, being $T\ped c$ the critical temperature. The CPRs have been normalized to the maximum value of the current at $0.05\;T\ped c$. The color-gradient in (a), (b) and (c) represents the temperature range for the $0$-state (light blue), the $\uppi$ state (light red), and the width of the $0-\uppi$ transition region (yellow region), obtained from the CPRs in (d), (e) and (f).}
	\label{IcT_sim_JJs}
\end{figure}

In Figure~\ref{IcT_sim_JJs} (a), \ref{IcT_sim_JJs}~(b) and \ref{IcT_sim_JJs}~(c), we show the comparison between the $I\ped c(T)$ curves measured down to $\SI{20}{\milli\K}$ at zero field for the junctions with \ce{GdN} barriers $d\ped F=\SI{3.0}{\nm}$, $\SI{3.5}{\nm}$ and $\SI{4.0}{\nm}$ (black points), respectively, and the simulations obtained with the tight-binding BdG lattice model (red straight lines). In the insets, we report the measured $I\ped c(T)$ values down to dilution temperatures. The experimental data evolve from a plateau over a wide range of temperatures (a few Kelvins) observed for the junctions with \ce{GdN} thickness $d\ped F=3.0$ and $\SI{3.5}{\nm}$ into a non-monotonic $I\ped c(T)$ curve for the junction with $d\ped F=\SI{4.0}{\nm}$. The agreement between numerical outcomes and experimental data is certified by the capability to reproduce the unconventional plateau (Figure~\ref{IcT_sim_JJs} (a) and \ref{IcT_sim_JJs}~(b)) and the non-monotonic behavior (Figure~\ref{IcT_sim_JJs} (c)). 

Details on Hamiltonian parameters used in the simulations can be found in Supplementary Note 2. However, we here briefly discuss the spirit of our lattice modelling. All the energy parameters are expressed in dimensionless units where the energy scale is the hopping $t$ in the FI.  The strenght of the SOC, $\alpha$, is scaled by $ta$ (with $a$ lattice constant), while the Josephson current is calculated in units of $J_{0}=\text{e}\Delta$. In our simulations, we fix  $t = 1$, $\mu\ped{FI} = 0$, $\mu\ped{s} = 3$, $\Delta = 0.005$, $h = 0.25$. Further, we note that \ce{NbN} (S leads) and \ce{GdN} (FI barrier) are characterized by almost equal hopping parameters~\cite{Litinskii1989,LeuenbergerGdN:article,Marsoner2015}, which are set equal $t\ped{s}=t$ for the sake of simplicity. The choice of assuming different chemical potentials for the S and FI regions is made in order to model the experimental devices as tunnel junctions with a ferromagnetic half-metallic \ce{GdN} barrier, as experimentally observed~\cite{Watcher1980} and predicted by full atomistic simulations~\cite{Duan2005,Larson2006}. The estimate for the exchange energy $h$ is chosen in agreement to the exchange field measured in several materials and is kept fixed to that of the bulk \ce{GdN}~\cite{Pal2015,Cascales2019,LeuenbergerGdN:article,Manchon2015,Marsoner2015}. This is consistent with $t\approx\SI{3}{\electronvolt}$ and the experimental constant lattice of \ce{GdN} is $a\ped{GdN}=\SI{4.974}{\angstrom}$~\cite{Litinskii1989,LeuenbergerGdN:article,Marsoner2015}.

When modelling the experiments, we use $\alpha$ as a measure of the spin-mixing and it is chosen to be $\alpha=0.04$, unless otherwise indicated. Although we choose a small spin-orbit field so that $\alpha\ll h$, it breaks the spin symmetry at interfaces and is sufficient to cause the generation of long-range equal triplet-correlation pairs with total spin projection $S_{z}=\pm1$.

\begin{figure}
	\centering
	\includegraphics[width=0.8\columnwidth]{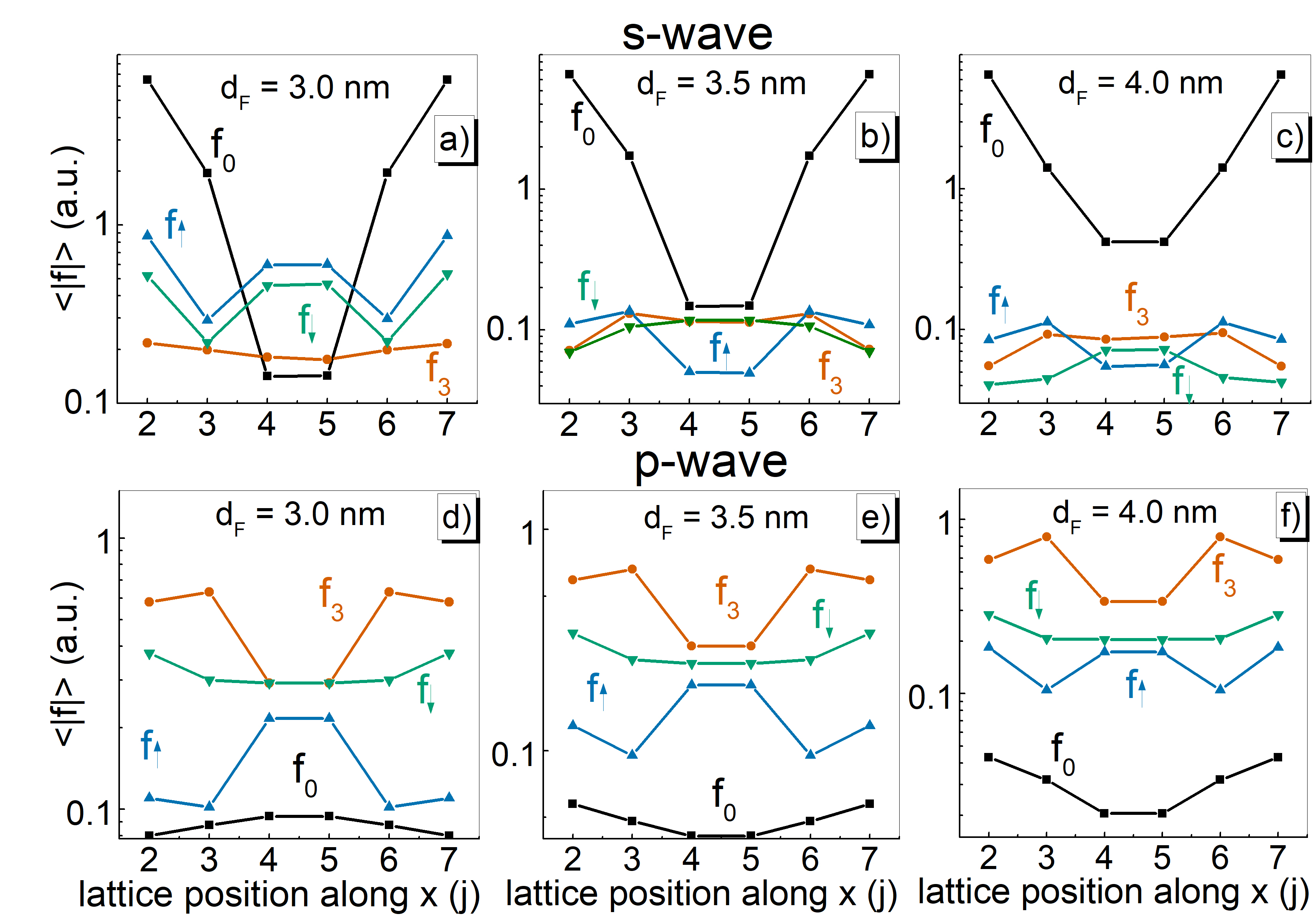}
	\caption[]{\textbf{S- and p-wave spin-singlet and triplet pair-correlation amplitudes in the ferromagnetic insulator barrier}. In (a), (b) and (c), the amplitudes of the ensamble average of the s-wave correlation functions $\left<\abs{f}\right>$, determined by numerical simulations at temperature $T=0.025\;T\ped c$, being $T\ped c$ the critical temperature, are shown as a function of the lattice position in the barrier along the $\mathbf{x}$ direction (with index $j$) for the junctions with \ce{GdN} thickness $d\ped F=3.0,\;3.5$ and $\SI{4.0}{\nm}$, respectively. $f_0$ is the spin-singlet (black line and square symbols), $f_3$ is the opposite-spin triplet (red line and circles) and $f_{\uparrow}$ ($f_{\downarrow}$) is the equal-spin triplet with up (down) $S_z$ projection (blue line and up-triangle symbols, and green line and down-triangle symbols, respectively). In (d), (e) and (f), we show the same correlation functions components for the p-wave symmetry.}
	\label{s_p_corrs}
\end{figure}

The numerical simulations in Figure\ref{IcT_sim_JJs} are performed on lattices characterized by: (a) $L=8$, $W=24$, (b) $L=8$, $W=28$, (c) $L=8, W=32$, expressed in units of lattice sites. Tunnel junctions experience an exponential suppression of the critical current when  increasing the barrier thickness~\cite{Barone1982}. In our model, this implies dealing with systems of few lattice sites, hence, we choose $L=8$ and keep it fixed in all the numerical simulations, in agreement with the short-junction limit. However, the main effect of increasing the experimental sample thickness (and so the magnetic area of the FI) consists in enhancing the magnetic activity of the junction~\cite{Senapati2011,Pal2014,Caruso2019,Ahmad2020JOSC}.  In our model, we manage to mimic this effect by changing the flux of the exchange field $\Phi(h) = LWh$ through the JJ (by the means of the width of the barrier $W$) and by tuning the impurity potential strength $V\ped {imp}$ (thus, changing the influence of disorder effects in the system). Therefore, we use these quantities as effective control parameters when modelling the peculiar behavior of the $I\ped{c}(T)$ curve in each experimental device. For the $I\ped {c}(T)$ simulations in Figure~\ref{IcT_sim_JJs} (as well as for the corresponding correlation functions in Figure~\ref{s_p_corrs}), we set $V\ped{imp} = 0.3$ for the simulated curve in (a), $V\ped{imp} = 0.37$ in (b), and $V\ped{imp} = 0.23$ in (c). Here, the presence of random on-site impurities requires the need to perform ensamble averages over several samples (see Supplementary Note 2 for details).

We notice that the Hamiltonian parameters, as well as the lattice size, have no microscopic (atomistic) origin and are chosen to describe the main mechanisms that are expected to occur in the experimental devices. Even though the lattice size is scaled down compared to the experimental system, we think that our theoretical model gives qualitatively an accordance with the experimental results as long as the model parameters are adjusted accordingly. 

We can relate the plateau in the $I\ped c(T)$ curve to an overall broadening of a $0-\uppi$ transition in temperature. The calculated CPRs in Figures~\ref{IcT_sim_JJs} (d), \ref{IcT_sim_JJs}~(e) and \ref{IcT_sim_JJs}~(f) indicate that at low temperatures the JJs are in the $0$-state (light blue gradient region in Figures~\ref{IcT_sim_JJs} (a), \ref{IcT_sim_JJs}~(b) and \ref{IcT_sim_JJs}~(c)), while at temperatures above $T=0.7\;T\ped c$ the JJs are in the $\uppi$ state (red gradient region). Compared to what has been theoretically and experimentally observed in $0-\uppi$ SFS and SFIS JJs~\cite{Ryazanov2001,Bannykh2009,Wild2010,Feofanov2010,Goldobin2013}, when the plateau is measured in the JJs with $d\ped F=\SI{3.0}{\nm}$ and $\SI{3.5}{\nm}$ in Figures~\ref{IcT_sim_JJs} (a) and \ref{IcT_sim_JJs}~(b), the CPRs exhibit the presence of higher order harmonics in the Josephson current $J$ for a wide range of temperatures (yellow gradient region). The transition region is reduced when the $I\ped c(T)$ curve gradually points towards a non monotonic behavior, as shown in Figure~\ref{IcT_sim_JJs} (c). In all the cases reported in Figure~\ref{IcT_sim_JJs}, the $0-\uppi$ transition extends over a few Kelvins in temperature around $\SI{4.2}{\K}$, in agreement with previous findings~\cite{Pal2014}. 

In Figure~\ref{s_p_corrs}, we show the amplitude of the correlation functions $\left<|f|\right>$ determined from numerical simulations for the three devices at $T = 0.025\;T\ped c$ (corresponding to $\SI{0.3}{\K}$) and $\phi=0$, where $\phi$ is the phase-difference across the device. The correlation functions are determined for the spin-singlet ($f_0$), spin-triplet with opposite spins ($f_3$) and equal-spin triplet functions ($f_{\uparrow}$ and $f_{\downarrow}$), both in s-wave (Figures~\ref{s_p_corrs} (a), \ref{s_p_corrs}~(b) and \ref{s_p_corrs}~(c)) and p-wave symmetries (Figures~\ref{s_p_corrs} (d), \ref{s_p_corrs}~(e) and \ref{s_p_corrs}~(f)), as a function of the position in the lattice along the $\mathbf{x}$ direction, with index $j=1,\dots,L$. In order to assure the total antisymmetry of the fermionic wave-function, triplet superconductivity for even-frequency pairing is conventionally of  p-wave type~\cite{Tanaka2007}. As shown in the following, for symmetry reasons here the dominant orbital part in the triplet pairing channel happens to be of s-wave type. We use $\left<\right>$ to indicate the ensemble average, due to the presence of random on-site impurities in the 2D-lattice. Details on the calculation of the spatial profile of the correlation function can be found in the Supplementary Note 1. All the cases show a dominant s-wave singlet component $f_{0}$ at the superconductor/barrier interface that strongly decays toward the middle of the barrier thickness. This is reasonable because the sides of the FI-layer are attached to the superconducting leads with a usual Bardeen-Cooper-Schrieffer s-wave symmetry~\cite{Ambegaokar1963} and, due to the proximity effect, the singlet pair wavefunction enters the barrier. In the middle of the barrier (lattice position $j=4$), where the spin mixing and the exchange field effects take place, a competition between the s-wave triplet and singlet pair amplitudes arises. On the contrary, for the p-wave case, the singlet component $f_0$ turns out to be much lower than the corresponding s-wave one. At the same time, we may observe a prevalence of the zero-spin p-wave triplet component $f_{3}$ at the superconductor/barrier interface, while in the middle of the barrier thickness the spin-aligned triplet correlations become relevant. These results are justified by symmetry considerations~\cite{Eschrig2011, Eschrig2015, Bergeret2005}. Indeed, for the s-wave symmetry, the singlet is an even-frequency function, while the triplets are odd-frequency. The viceversa is valid for the p-wave case. 

The $\SI{3.0}{\nm}$-thick barrier junction exhibits s-wave triplet correlations functions larger than the singlet one, with a major contribution provided by the equal-spin triplet component with $S_z=+1$, $f_{\uparrow}$  (Figure~\ref{s_p_corrs} (a)). For what concerns the p-wave spin-correlation functions for this device, $f_{3}$ provides the main contribution at the borders, while $f_{\downarrow}$ competes with $f_3$ in the middle of the barrier (lattice position $j=4$), as shown in Figure~\ref{s_p_corrs} (d). Moreover, the opposite- and equal-spin p-wave triplet components are nearly a factor $2$ larger than the corresponding s-wave singlet component. By increasing the thickness of the barrier, thus gradually pointing towards an incipient $0-\uppi$ transition with a non-monotonic behavior in the $I\ped c(T)$ curve, in the s-wave cases we can observe a progressive suppression of the equal-spin triplet components and a dominant spin-singlet channel. At the same time, in the p-wave case, we observe a slight reduction of the ratio between the equal-spin triplets ($f_{\uparrow}$, $f_{\downarrow}$) and the major zero-spin component ($f_{3}$). Thus, the p-wave opposite spin-triplet components are of the same order of magnitude compared to the corresponding s-wave spin-singlet component, while the equal-spin triplet components are instead reduced.

\begin{figure}
	\centering
	\includegraphics[width=0.8\columnwidth]{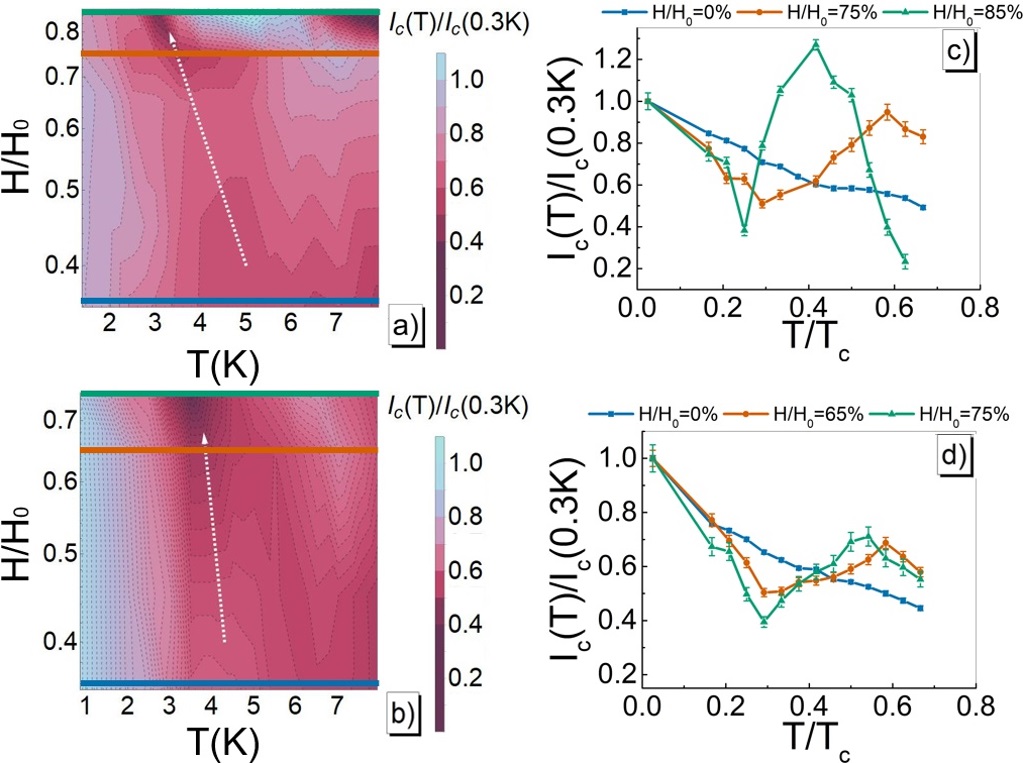}
	\caption[]{\textbf{Tuning of the temperature behavior of the critical current in presence of an external magnetic field.} Normalized critical current $I\ped c(T,H/H_0)/I\ped c(\SI{0.3}{\K},H/H_0)$ density plots as a function of the percentage of magnetic field periodicity $H/H_0$ and the temperature $T$, for the Josephson junctions with \ce{GdN} thickness (a) $d\ped F=\SI{3.0}{\nm}$ and (b) $d\ped F=\SI{3.5}{\nm}$. The critical current $I\ped c$ values at each temperature $T$ are measured by fixing the external magnetic field to $H/H_0$, where $H_0$ is the amplitude of the first-lobe of the Fraunhofer pattern measured at the same temperature $T$. More details can be found in Methods - Experimental $I\ped c(T)$ curves at zero- and finite-field. Blue, red and green lines refer to the cross sections reported in (c) and (d): blue squares for $H/H_0=0\%$, red circles for $H/H_0=75\%$ in (c) and $H/H_0=65\%$ in (d), green triangles for $H/H_0=85\%$ in (c) and $H/H_0=75\%$ in (d). Straight lines in plots (c) and (d) are only a guide for the eye. The error bar on each measured point is of the order of few percents and it is due to thermally-induced $I\ped c$ fluctuations\cite{Massarotti2015}. The white dashed arrows in (a) and (b) are a guide for the eye and highlight the shift of the minimum in the $I\ped c(T,H/H_0)/I\ped c(\SI{0.3}{\K},H/H_0)$ by increasing $H/H_0$.}
	\label{Fig4:29}
\end{figure}

In order to investigate the peculiar transport properties arising in these systems, we have probed the $I\ped{c}(T)$ response to an external magnetic field applied in the plane of the JJs. In Figure~\ref{Fig4:29} we show the evolution of the normalized critical current $I\ped c(T,H/H_0)/I\ped c(\SI{0.3}{\K},H/H_0)$ as a function of a weak magnetic field $H/H_0$, where $H_0$ is the amplitude of the first lobe of the Fraunhofer pattern curve, acquired by applying the magnetic field from $+\SI{2.4}{\milli\tesla}$ to $-\SI{2.4}{\milli\tesla}$. $H_0$ is estimated at each investigated temperature $T$ (from $T=\SI{0.3}{\K}$ to $T=\SI{8}{\K}$). Details on the measurement procedure can be found in the Methods section - Experimental $I\ped c(T)$ curves at zero- and finite-field. 

The results are reported in Figure~\ref{Fig4:29} in the two density-plots \ref{Fig4:29}~(a) and \ref{Fig4:29}~(b) for the junctions with $d\ped F=\SI{3.0}{\nm}$ and $\SI{3.5}{\nm}$, respectively. Increasing the field $H/H_0$, the plateau structure at zero field evolves into a non-monotonic behavior with a minimum (dark region around $70-80\%H_0$ and between $2$ and $\SI{4}{\K}$) and a maximum (bright region around $70-80\%H_0$ and between $4$ and $\SI{6}{\K}$). The effect is more pronounced for the JJ with $d\ped F=\SI{3.0}{\nm}$. The blue, green and red dashed line cuts are related to the cross-section curves reported in Figure~\ref{Fig4:29} (c) and \ref{Fig4:29}~(d), where the gradual appearance of an enhanced dip and a non-monotonic behavior in the normalized $I\ped{c}(T)$ curves can be observed by increasing $H/H_0$. The dependence of the $I\ped c$ as a function of the normalized magnetic field $H/H_0$ is reported in Supplementary Figure 2 for the JJ with \ce{GdN} thickness $d\ped F=\SI{3.0}{\nm}$  at three selected temperatures: $\SI{0.3}{\K}$, $\SI{3}{\K}$ where a minimum of the $I\ped c(T)$ curve is measured for $H/H_0=75\%$, and $\SI{7}{\K}$ where a maximum of the $I\ped c(T)$ is observed for the same value of $H/H_0$. While a standard Fraunhofer-like $I\ped c(H)$ dependence is recovered at the three selected temperatures, for magnetic fields close to a quantum flux and at high temperature, e.g. $\SI{7}{\K}$, $I\ped c$ is larger than the value measured at low temperature, e.g. $\SI{0.3}{\K}$. In order to stress this point, the IV curves measured at $H/H_0=75\%$ for the three selected temperatures are shown in Supplementary Figure 2 as a term of reference.
	
Such a progressive variation of the $I\ped c(T)$ curves in presence of an external magnetic field is not observed in junctions with thinner \ce{GdN} barriers, as shown in the Supplementary Figure 3, in which we report the $I\ped c(T,H/H_0)/I\ped c(\SI{0.3}{\K},H/H_0)$ density-plot measured on a \ce{NbN}-\ce{GdN}-\ce{NbN} junction with \ce{GdN} barrier thickness $d\ped F=\SI{1.5}{\nm}$. For this device, a standard Ambegaokar-Baratoff (AB) trend~\cite{Ambegaokar1963,ErratumAmbegaokar1963} for the $I\ped c(T)$ curve is preserved in presence of the external magnetic field. Compared to devices with thin \ce{GdN} barriers, the samples analyzed in this work are indeed sensitive to a weak magnetic field~\cite{Pal2014,Massarotti2015,Caruso2019}. A finite shift of the order of $\SI{0.1}{\milli\tesla}$ in the Fraunhofer pattern curves arises when ramping the field from positive to negative values, and viceversa~\cite{Blamire2013}. Even if the strength of the external magnetic field is not enough to generate a complete magnetic ordering, slight modifications in the microscopic structure of the barrier arise~\cite{Cullity2011}, which has been already predicted to occurr in systems with tunable domain walls~\cite{Baker2014}, intrinsic SOC~\cite{Liu2010} and magnetic impurities~\cite{Pal2018}. At zero field, the magnetic disorder is maximum and likely introduces electronic defect states in the barrier~\cite{Maity2020}. As the field increases, the system undergoes towards a more ordered phase, and hence defect states density reduces. Therefore, the tunability of the $I\ped c(T)$ shape from the plateau towards a non-monotonic curve by applying an external magnetic field can be related to a reduction of the disorder in the barrier.

\begin{figure}
	\centering
	\includegraphics[width=0.8\columnwidth]{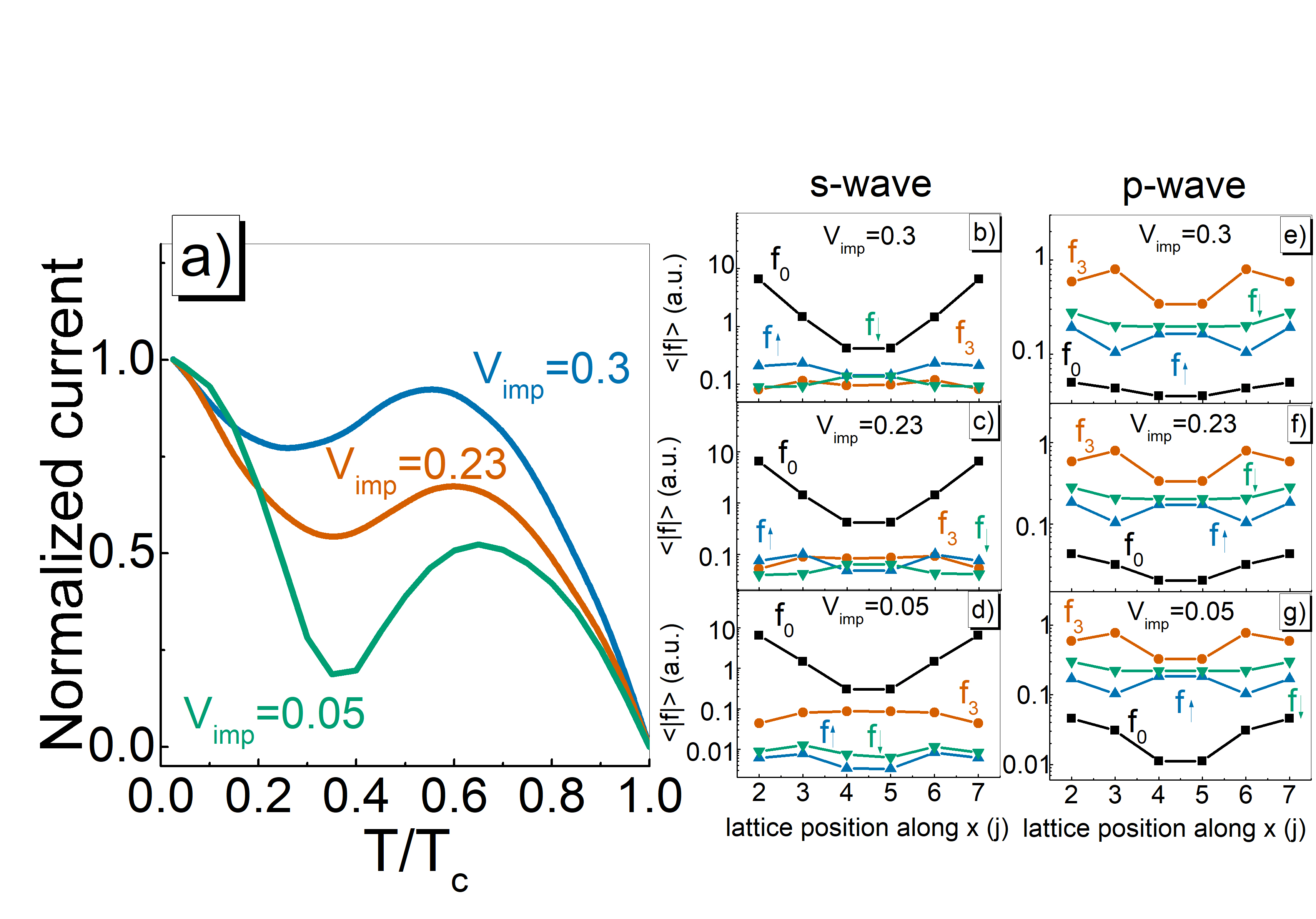}
	\caption[]{\textbf{Simulated temperature behavior of the critical current and calculated pair-correlation amplitudes as a function of the impurity potential.} In (a), normalized critical current $I\ped c$ vs. temperature $T$ curves simulated with the two-dimensional lattice model at fixed dimensions for three different impurity potential $V\ped{imp}$ values: $V\ped{imp} = 0.05$ (green curve), $V\ped{imp} = 0.23$ (red curve) and $V\ped{imp} = 0.3$ (blue curve). Simulation parameters are reported in the Supplementary Note 2. The current is normalized to the maximum of the current-phase relation at the lowest investigated temperature $T$, while $T$ is normalized to the critical temperature $T\ped c$. In (b), (c) and (d), calculated s-wave ensamble average of the pair amplitude $\left<\abs{f}\right>$ in arbitrary units for different impurity potential values $V\ped{imp}$ in (a). In (e), (f) and (g), calculated p-wave ensamble average of the pair amplitude $\left<\abs{f}\right>$ in arbitrary units for different impurity potential values $V\ped{imp}$ in (a). $f_0$ is the spin-singlet component (black line and square symbols), $f_3$ is the opposite-spin triplet component (red line and circle symbols),  $f_{\uparrow(\downarrow)}$ is the up (down) equal-spin triplet component (blue lines and up-triangle symbols, and green lines and down-triangle symbols, respectively). Both s- and p-wave data are reported on a log-scale.}
	\label{Vimp_vs_Field}
\end{figure}

This picture is supported by numerical simulations obtained when changing the strength of the impurity potential in the 2D-lattice model while keeping fixed all the other parameters. As a matter of fact, in order to have a good agreement with experimental data, we model the \ce{GdN} as a ferromagnetic half metal~\cite{Watcher1980,Duan2005,Larson2006}. Hence, local impurity potentials in the FI barrier are assumed to induce small site-dependent fluctuations of the chemical potential. In our approach, the coexistence of spin mixing mechanisms, promoted by SOC-like interactions, and on-site impurities model the magnetic disorder. In Figure~\ref{Vimp_vs_Field} (a), in fact, we can notice that the characteristic $0-\uppi$ behavior is modified by increasing the impurity potential $V\ped{imp}$. The enhancement of the impurity strength produces a shift of the minimum of the curve towards lower temperatures and higher critical current values, with a consequent broadening of the typical $0-\uppi$ cusp that progressively gives rise to the plateau. Viceversa, decreasing $V\ped{imp}$, one can recover the $0-\uppi$ transition. Details on the parameters are reported in the Supplementary Note 2.

In Figure~\ref{Vimp_vs_Field}, we finally report the s- and p-wave correlation functions corresponding to simulated $I\ped{c}(T)$ curves for different impurity potentials $V\ped{imp}$ in Figure~\ref{Vimp_vs_Field} (a): $V\ped{imp} = 0.3$, $V\ped{imp} = 0.23$ and $V\ped{imp} = 0.05$. We here take as a reference the JJ with $d\ped F=\SI{4.0}{\nm}$, i.\,e. the simulations for lattice dimensions $L=8,\;W=32$. For the s-wave symmetry reported in Figures~\ref{Vimp_vs_Field} (b), \ref{Vimp_vs_Field}~(c) and \ref{Vimp_vs_Field}~(d), the effect of increasing the impurity strength $V\ped{imp}$ results in a pronounced enhancement of the equal-spin triplet pairing correlations, $f_{\uparrow}$ and $f_{\downarrow}$, while the p-wave components appear to be approximately unaffected by disorder (e.g. Figures~\ref{Vimp_vs_Field} (e), \ref{Vimp_vs_Field}~(f) and \ref{Vimp_vs_Field}~(g)).

\section{Discussion}
\label{discussion}

The theoretical results in Figures~\ref{IcT_sim_JJs}-\ref{s_p_corrs} show that the characteristic behavior of the $I\ped{c}(T)$ is related to the amplitude of the different s-wave spin-correlation functions. In Table~\ref{Tab:corr}, we summarize the values of the pair-correlations in the middle of the barrier thickness (lattice position $j=4$), in units of the majority zero-spin component, i.\,e. $f_{0}$ for the s-wave (Table~\ref{Tab:corr} A) and $f_{3}$ for the p-wave cases (Table~\ref{Tab:corr} B), respectively. Indeed, we observe a general decrease in the relative weight of the s-wave equal-spin triplet components ($f_{\uparrow}$ and $f_{\downarrow}$) in the junctions that show an increasing non-monotonicity of the $I\ped{c}(T)$ curves. Hence, the more $I\ped{c}(T)$ exhibits a behavior approaching the $ 0- \uppi $ regime, the lower is the weight of the s-wave equal-spin correlations. This is in agreement with the fact that spin-aligned supercurrents are insensitive to exchange field and, thus, cannot give rise to $0-\uppi$ transitions.

\begin{figure}
	\centering
	\includegraphics[width=0.8\columnwidth]{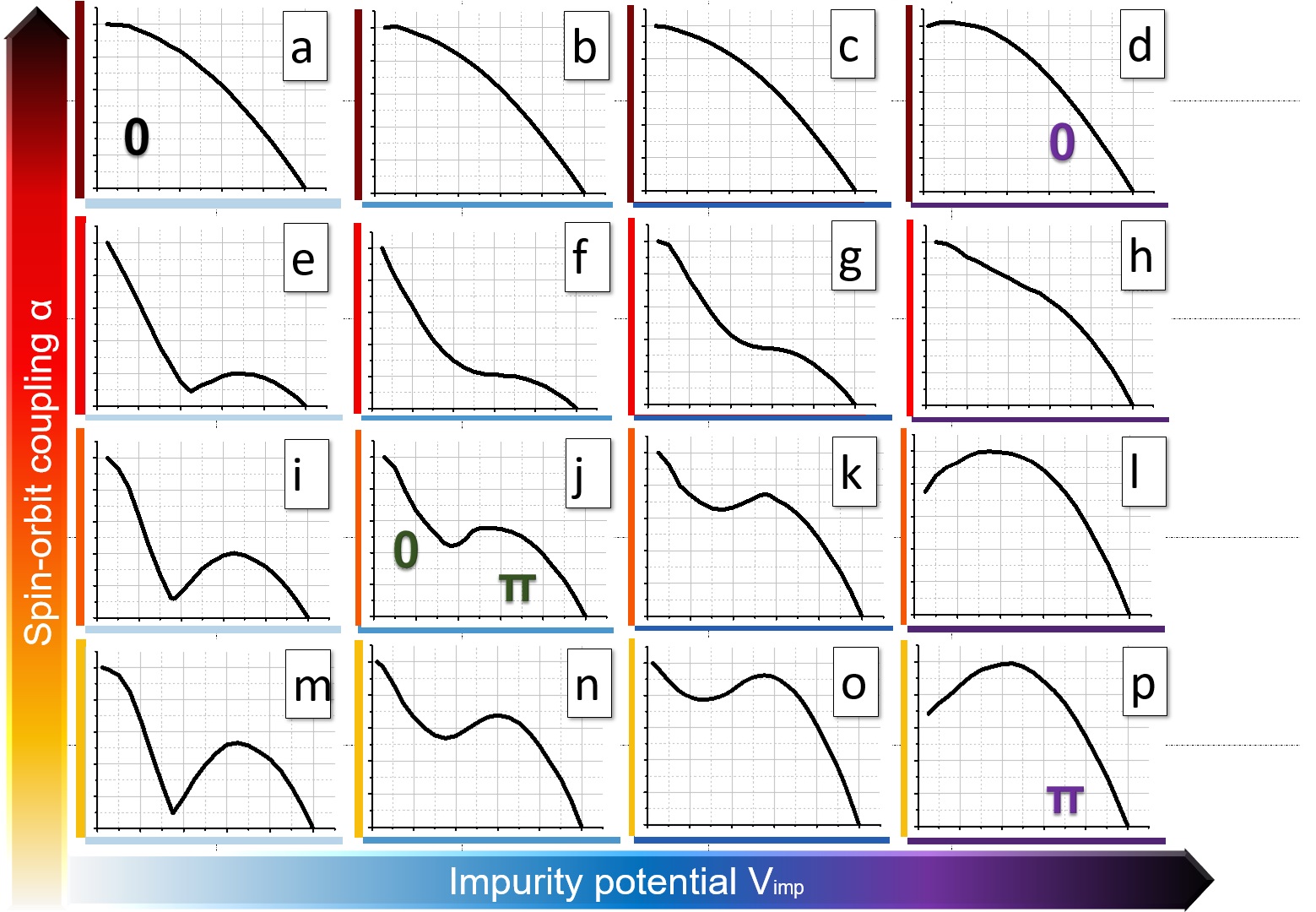}
	\caption[]{\textbf{Competition between the spin-orbit coupling and the impurity potential and their effect on the temperature behavior of the critical current.} Normalized critical current $I\ped{c}$ vs. temperature $T$ curves simulated with the tight-binding Bogoliubov-De Gennes two-dimensional lattice model as a function of the spin-orbit coupling $\alpha$ and the on-site impurity potential $V\ped{imp}$. In (a) $\alpha = 0.2$, $V\ped {imp} = 0.05$; (b) $\alpha = 0.2$, $V\ped {imp} =0.23$; (c) $\alpha = 0.2$, $V\ped {imp} = 0.3$; (d) $\alpha = 0.2$, $V\ped {imp} = 0.5$; (e) $\alpha = 0.1$, $V\ped {imp} = 0.05$; (f) $\alpha = 0.1$, $V\ped {imp} =0.23$; (g) $\alpha = 0.1$, $V\ped {imp} = 0.3$; (h) $\alpha = 0.1$, $V\ped {imp} = 0.5$; (i) $\alpha = 0.07$, $V\ped {imp} = 0.05$; (j) $\alpha = 0.07$, $V\ped {imp} =0.23$; (k) $\alpha = 0.07$, $V\ped {imp} = 0.3$; (l) $\alpha = 0.07$, $V\ped {imp} = 0.5$; (m) $\alpha = 0.04$, $V\ped {imp} = 0.05$; (n) $\alpha = 0.04$, $V\ped {imp} =0.23$; (o) $\alpha = 0.04$, $V\ped {imp} = 0.3$; and (p) $\alpha = 0.04$, $V\ped {imp} = 0.5$. In all the panels, the $I\ped c$ ($\mathbf{y}$-axis) is normalized to its value at the lowest temperature, i.\,e. $T=\SI{300}{\milli\K}$, while the $T$ ($x$-axis) is normalized to the critical temperature of the device $T\ped c$. Red-color scale refers to increasing values of $\alpha$, while blue-color scale refers to increasing $V\ped{imp}$ values, with parameters reported in Discussion and in the Supplementary Note 2. The scale on the $y$-axis on each plot ranges from $0$ to $1.1$, as on the $\mathbf{x}$-axis. Minor thicks represent an increment of $0.1$. We also highlight in panels (a), (d), (j) and (p) the state of the Josephson junction: $0$, $0-\uppi$ or $\uppi$.}
	\label{diagram}
\end{figure}

In Figure~\ref{diagram}, we show how the impurities and the SOC affect the $I\ped c(T)$ shape. Lattice dimensions are $L=8, W=32$, i.\,e. they refer to the JJ with $d\ped F=\SI{4.0}{\nm}$. To accomplish the $I\ped {c}(T)$ diagram, we select the values for $\alpha$ and $V\ped{imp}$ as described in Fig.~\ref{diagram} (a-p). For small values of $\alpha$ and $V\ped{imp}$ (bright red- and blue-scales), the simulated $I\ped c(T)$ curve shows a cusp-like $0-\uppi$ transition, provided that the exchange field $h$ in the junction is non-zero, as it occurs in SFS JJs tipically reported in literature~\cite{Ryazanov2001,Bannykh2009,Wild2010,Feofanov2010,Goldobin2013}. By increasing $\alpha$ (dark red-scale), the main effect is to reduce the height of the second maximum in the $I\ped c(T)$ curve, without recovering the plateau structure observed in SFIS JJs. At very large $\alpha$ (see panel (a) in Figure\ref{diagram}), the $0-\uppi$ transition is washed out and an AB-like shape sets in, stabilizing a "$0$"-phase. In this case the main contribution is expected from the spin-singlet, though the spin-triplet correlations are increased compared to the cases with smaller $\alpha$. 

At the same time, by keeping the spin-orbit field weak and by increasing $V\ped{imp}$ (dark blue-scale), the minimum of the $0-\uppi$ transition occurs at higher critical current values and it is broadened in temperature, but always showing a non-monotonic trend for the $I\ped c(T)$. The characteristic plateau structure is observed only when considering a combined effect of SOC and impurities, once fixed the dimensions of the system. As it is shown for the SFIS JJ with $d\ped F=\SI{3.0}{\nm}$ in Figure~\ref{IcT_sim_JJs} (a) and Figure~\ref{s_p_corrs} (a), the formation of the plateau goes along with the coexistence of comparable spin-singlet and triplet superconductivity. In the limit of large $V\ped {imp}$ and $\alpha$ (see panel (d) in Figure\ref{diagram}), an AB-like behavior is recovered. This latter corresponds to a stable "$0$"-phase, reflecting the fact that, in the competition between SOC and impurity scattering, the equilibrium state is dominated by $\alpha$. This also confirms the presence of a threshold value of $\alpha$ (at fixed value of $h$), above which the JJ is always in the "$0$"-phase~\cite{Asano2019,minutillo2021realization}. In the limit of large $V\ped{imp}$ and small $\alpha$ (see panel (p) in Figure\ref{diagram}), the $0-\uppi$ transition is shifted towards very low $T$ values, stabilizing a "$\uppi$"-phase almost over the whole temperature range. This evidence is given by the sharp decrease of $I\ped c$ when the temperature drops. In this regime, in agreement with Figure~\ref{Vimp_vs_Field}, we predict an enhanced contribution of the s-wave spin-triplet components due to the interplay of spin-orbit and disorder. 

A transition between the peculiar plateau-shape of the $I\ped c(T)$ curve towards an incipient $0-\uppi$ curve is experimentally observed increasing the strength of an external weak magnetic field (Figure~\ref{Fig4:29}). The position in temperature of the $I\ped c(T)$ dip is an important benchmark relating the $0$-$\uppi$ transition induced by the weak magnetic field to the combined effect of impurities, exchange field fluctuations and spin-orbit coupling in the simulations. For weak on-site impurity potential, by increasing $\alpha$, the minimum of the $I\ped c(T)$ curve occurs at the same temperature. Instead, as shown in Figure\ref{Vimp_vs_Field}, when increasing $V\ped{imp}$, the minimum is shifted in temperature, as it occurs in the experimental $I\ped c(T)$ curves at a finite external magnetic field. We highlight the evolution of the $0$-$\uppi$ transition and the dip shift in the experimental data in Figure~\ref{Fig4:29} (a) and \ref{Fig4:29}~(b) by using the dashed white arrow in figure, which is here only a guide for the eye.

In conclusion, in this work we have investigated on the occurence of the unconventional $I\ped{c}(T)$ behaviors observed in SFIS JJs. The presence of a plateau extended over a wide range of temperatures and the peculiar non-monotonic behavior in the $I\ped{c}(T)$ when increasing the thickness of the barrier can be explained in terms of the coexistence of spin-singlet and triplet superconductivity, whose correlation functions have been calculated by using a tight-binding BdG descripion of the system~\cite{Asano2001, Asano2019, Furusaki1994}. This approach highlighted also the role played by the disorder in the barrier. At the same time, the presence of a spin-mixing effect, in this context provided by the spin-orbit interaction, is crucial to reproduce the characteristic plateau in the $I\ped{c}(T)$ curves. Furthermore, the obtained results confirm that the reinstatement of an overall ordering in the system by the means of an external magnetic field points towards the recovery of more standard $0-\uppi$ transition. Within this picture, the application of a weak magnetic field represents a tool for controlling the relative weight of equal-spin-triplet transport in SFIS JJs. 

\section{Methods}

\subsection*{Experimental $I\ped c(T)$ curves at zero- and finite-field}
\label{Methods_A}

The $I\ped c(T)$ measurements at zero-field have been performed by using an evaporation cryostat from $\SI{0.3}{\K}$ up to the critical temperature of the devices $T\ped c\sim\SI{12}{\K}$~\cite{Caruso2018,Ahmad2020}. Measurements below $\SI{0.3}{\K}$ have been performed by using a wet dilution refrigerator Kelvinox400MX~\cite{Ahmad2020,Longobardi2012} and a dry dilution cryostat Triton400. The three refrigerators employed are equipped with customized low-noise filters anchored at different temperature stages~\cite{Caruso2018,Ahmad2020}. 

The $I\ped c(T)$ curves at finite external magnetic field have been measured by using the evaporation cryostat, which is equipped with a \ce{NbTi} coil able to generate a magnetic field perpendicular to the transport direction~\cite{	Caruso2018,Caruso2019}. The protocol followed is based on the measurement of the Fraunhofer pattern curve at fixed temperature for the spin-filter JJs with $d\ped F=\SI{3.0}{\nm}$ and $\SI{3.5}{\nm}$. The $I\ped c(H)$ curve is acquired ramping the magnetic field from positive to negative values, in a maximum field range of $\pm\SI{2.4}{\milli\tesla}$. The investigated range of temperatures corresponds to the plateau regime at zero field (from $\SI{2}{\K}$ to $\SI{8}{\K}$). Measurements at $\SI{0.3}{\K}$ have been performed as a term of comparison with the $I\ped c(T)$ curves at zero-field. 

At each temperature, we have measured the amplitude of the first lobe of the Fraunhofer pattern $H_0$. The error on $H_0$ is $3\%$. Then, we have estimated the critical current $I\ped c$ at different percentage of $H_0$, from $0\%H_0$ to $85\%H_0$ for the JJ with $d\ped F=\SI{3.0}{\nm}$. The $I\ped c$ for the JJ with $d\ped F=\SI{3.5}{\nm}$ at high temperatures is less than some nanoamperes~\cite{Caruso2019}, hard testing the observation of the Fraunhofer modulation. This limited the maximum magnetic field range investigated to $75\%H_0$. The step in field was chosen to be sufficiently small to avoid the interpolation of the $I\ped c$ from the pattern curve. Thus, the $I\ped c$ has been measured directly from the $I(V)$ curves at the field corresponding to $H/H_0$.  Moreover, to distinguish a peak structure in temperature, the step in temperature was fixed to $\SI{0.5}{\K}$. 

Given the small range of the field explored, the shift of the maximum in the Fraunhofer pattern due to the hysteretic magnetization of the barrier was of the order of  $\sim\SI{0.1}{\milli\tesla}$~\cite{Pal2014,Caruso2019}. We have removed the hysteretic shift in post-processing to guarantee that any measured effect is only related to the local magnetic field. The error on $I\ped c$ ranges from $1\%$ to $5\%$ for currents below the nanoamperes~\cite{Caruso2019,Ahmad2020}.

\section*{Data Availability}

The data that support the findings of this study are available from the corresponding authors upon reasonable request.

\section*{Code Availability}

Codes written for and used in this study are available from the corresponding authors upon reasonable request.

\begin{acknowledgments}
H.G.A., D.M. and F.T. acknowledge financial support from Università degli Studi di Napoli Federico II through the project: EffQul- Efficient integration of hybrid quantum devices - RICERCA DI ATENEO\_LINEA A, CUP: E59C20001010005. H.G.A., D.M. and F.T. also thank NANOCOHYBRI project (COST Action CA 16218).
\end{acknowledgments}

\section*{Author contributions}
	H.G.A and M.M. contributed equally. H.G.A., D.M. and F.T. conceived the experiments; A.P. and M.G.B. designed and realized the junctions; H.G.A., D.M. and R.Caruso carried out the measurements; M.M., R.Capecelatro, H.G.A. and R.Caruso worked on the data analysis; G.P., M.M., R.Capecelatro, and P.L worked on the theoretycal model code; H.G.A., M.M., R.Capecelatro, D.M., P.L., F.T., M.G.B. co-wrote the paper. All authors discussed the results and commented the manuscript.
	
\section*{Competiting interests}

All authors declare no competing financial and non-financial interests.


%

\clearpage
	
\clearpage
\clearpage
\begin{table}[t]
	\centering
	\caption[]{\textbf{S- and p-wave symmetry spin-correlations.} Ensamble average of the pair-correlation amplitudes $\left<\abs{f}\right>$, here represented as $f$, in the middle of the barrier (lattice position along $\mathbf{x}$ $j=4$) for Josephson junctions with \ce{GdN} thickness $d\ped F$: $f_{\uparrow}$ and $f_{\downarrow}$ for up- and down spin triplet correlation functions, respectively, $f_3$ for opposite-spin triplet, $f_0$ for spin-singlet. In (A) and (B), $f$ is reported for both s- and p-wave symmetry in units of the major zero-spin component: spin-singlet $f_{0}$ for the s-wave correlations and the zero-spin triplet $f_{3}$ for the p-wave correlations, respectively.}
	\begin{tabular}{lllllllllll}
		\toprule
		(A)& s-wave & & & (B)& & &&  p-wave & & \\
		\midrule
		$d\ped F$ ($\si{\nm}$) & $f_{\uparrow}/f_0$ & $f_{\downarrow}/f_0$ & $f_3/f_0$ & & & && $f_{\uparrow}/f_3$ & $f_{\downarrow}/f_3$ & $f_0/f_3$ \\
		\midrule
		$3.0$ 				   & $4.21$ 			& $3.23$ 			   & $1.28$    & & & && $0.74$ 			  & $1.00$ 				 & $0.32$\\	
		$3.5$ 				   & $0.34$ 			& $0.80$ 			   & $0.78$    & & & && $0.67$ 			  & $0.84$ 				 & $0.14$ \\
		$4.0$ 				   & $0.11$ 			& $0.15$ 			   & $0.20$    & & & && $0.51$ 			  & $0.60$ 				 & $0.06$  \\
		\bottomrule
	\end{tabular}\\
	\label{Tab:corr}
\end{table}

\end{document}